\documentclass[a4paper,10pt]{article}

\usepackage[english]{babel} 
\usepackage[latin2]{inputenc} 
\usepackage{graphicx} 
\usepackage{multirow} 
\usepackage[center]{caption}
\usepackage{geometry}
\usepackage{indentfirst}
\usepackage{txfonts}

\newenvironment{changemargin}[2]{%
  \begin{list}{}{%
    \setlength{\topsep}{0pt}%
      \setlength{\leftmargin}{#1}%
      \setlength{\rightmargin}{#2}%
      \setlength{\listparindent}{\parindent}%
      \setlength{\itemindent}{\parindent}%
      \setlength{\parsep}{\parskip}%
  }%
  \item[]}{\end{list}}

\usepackage{multirow}
\usepackage{epstopdf}

\title{Estimations of non-linearities in structural vibrations of string musical instruments}
\author{Kerem Ege \\
   Laboratoire Vibrations Acoustique,\\
   INSA-Lyon,\\
   25 bis avenue Jean Capelle,\\
   F-69621 Villeurbanne Cedex, France\\
   \texttt{kerem.ege@insa-lyon.fr}
   \and Marc Rébillat \\
   Département d'études cognitives,\\
   École Normale Supérieure Paris,\\
   29 Rue d'Ulm,\\
   75230 Paris, France\\
      \and Xavier Boutillon \\
   Laboratoire de Mécanique des Solides,\\
   École Polytechnique,\\
   91128 Palaiseau, France}
\date{}

\begin{document}


\twocolumn[
\maketitle
\begin{@twocolumnfalse}
\begin{changemargin}{1cm}{1cm}
Under the excitation of strings, the wooden structure of string instruments is generally assumed to undergo linear vibrations. As an alternative to the direct measurement of the distortion rate at several vibration levels and frequencies, we characterise weak non-linearities by a signal-model approach based on cascade of Hammerstein models. In this approach, in a chain of two non-linear systems, two measurements are sufficient to estimate the non-linear contribution of the second (sub-)system which cannot be directly linearly driven, as a function of the exciting frequency. The experiment consists in exciting the instrument acoustically. The linear and non-linear contributions to the response of (a) the loudspeaker coupled to the room, (b) the instrument can be separated. Some methodological issues will be discussed. Findings pertaining to several instruments - one piano, two guitars, one violin - will be presented.
\end{changemargin}\vspace{24pt}
\end{@twocolumnfalse}]

\section{Introduction: a linear behavior?}
Non-linear phenomena such as jump phenomenon, hysteresis or internal resonance appear when the transverse vibration of a bi-dimensional structure exceeds amplitudes in the order of magnitude of its thickness~\cite{TOU2002}. 

\begin{figure}[ht!]
\begin{center}
\includegraphics[width= 0.44\textwidth]{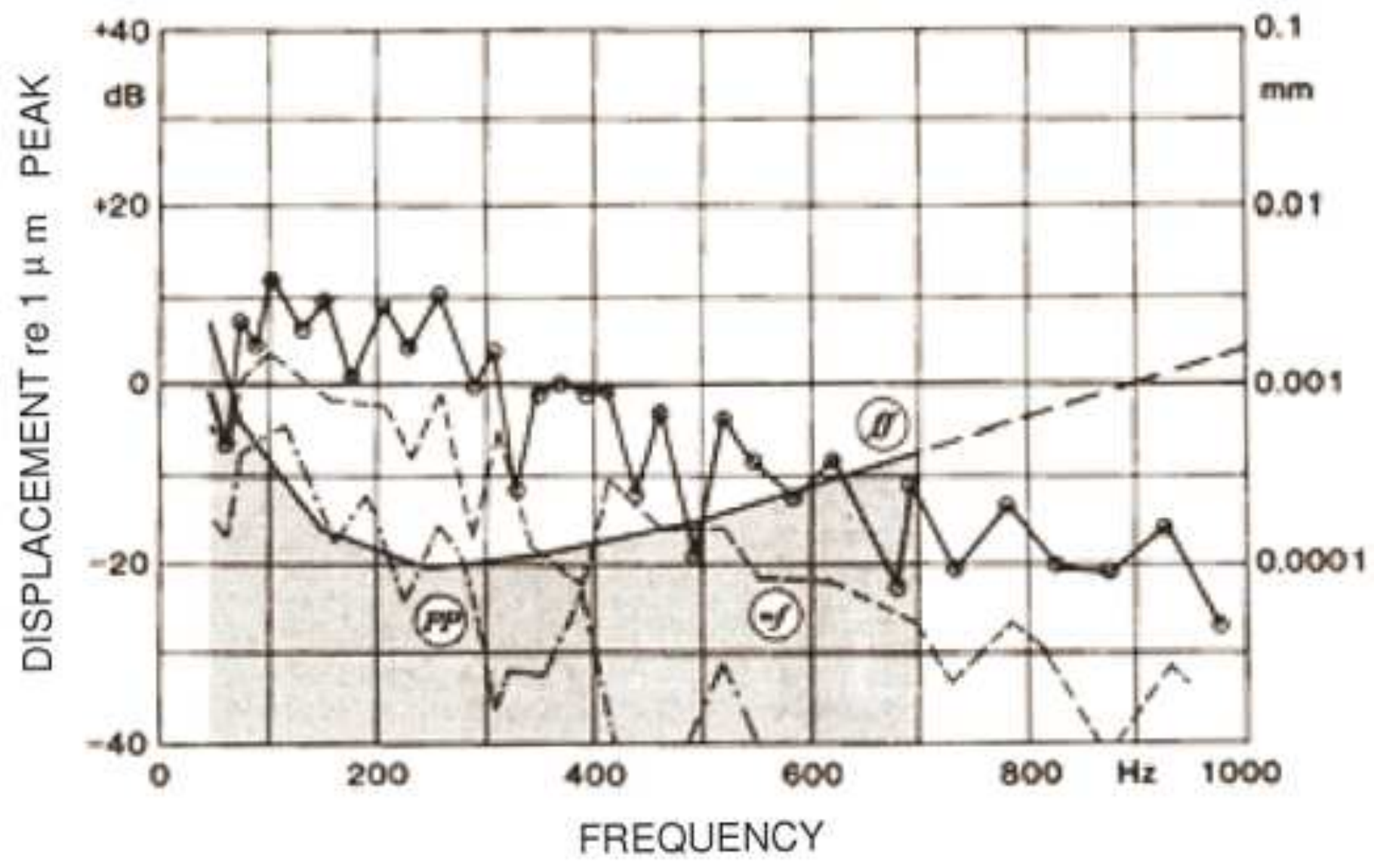}
\end{center}
\caption[Ask]{Vibration levels at the bridge of a grand piano when played \emph{pp} (dash-dotted line), \emph{mf} (dotted line) and \emph{ff} (solid line with $\bullet$ marks) for the notes $\mathbf{C_2}$ to $\mathbf{B_5}$ (fundamental frequencies $\approx$~60 to 950~Hz), according to~\cite{ASK1992}.}
\label{fig:Ask_soundb_disp}
\end{figure}

For string musical instrument (violin family, guitars, pianos...), the soundboard is generally assumed to undergo linear vibrations: the transverse motion $w$ remains in a smaller range than the board thickness. For example in the case of the piano, $w$ remains in a smaller range, even when the piano is played \emph{ff} in the lower side of the keyboard. Askenfelt and Jansson~\cite{ASK1992} report maximum values of the displacement at the bridge $w_{max}\approx6\cdot 10^{-6}$~m in the frequency range 80--300~Hz (Fig.~\ref{fig:Ask_soundb_disp}). This maximum value is less than $10^{-3}$ times the board thickness (around 8~mm). It can therefore be assumed that, to a high level of approximation, the vibration of the soundboard is linear.
 
The purpose of this article is thus to quantify experimentally such linear approximation. An original vibro-acoustical method is presented in Sec.~\ref{sec:chain} to isolate the soundboard non-linearity from that of the exciting device and to measure it. Soundboard intrinsic non-linearities of one upright piano, two guitars and one violin are then quantified using this method and results are presented in Sec.~\ref{sec:results}.

\section{A chain of two non-linear systems}
\label{sec:chain}

When dealing with non-linearities,  the non-linear contribution of the system under study (here a soundboard) has to be quantified, and isolated from the non-linear contribution of the exciting device, which can be an electromechanical exciter or electrodynamic loudspeaker for example. The vibro-acoustic method presented in this section faces this problem and allows the estimation of the non-linear contribution of the soundboard of the instruments in cases where it cannot be directly linearly driven.

\subsection{Notations}

\begin{figure}[h!]
\begin{center}
\includegraphics[width = 0.42\textwidth]{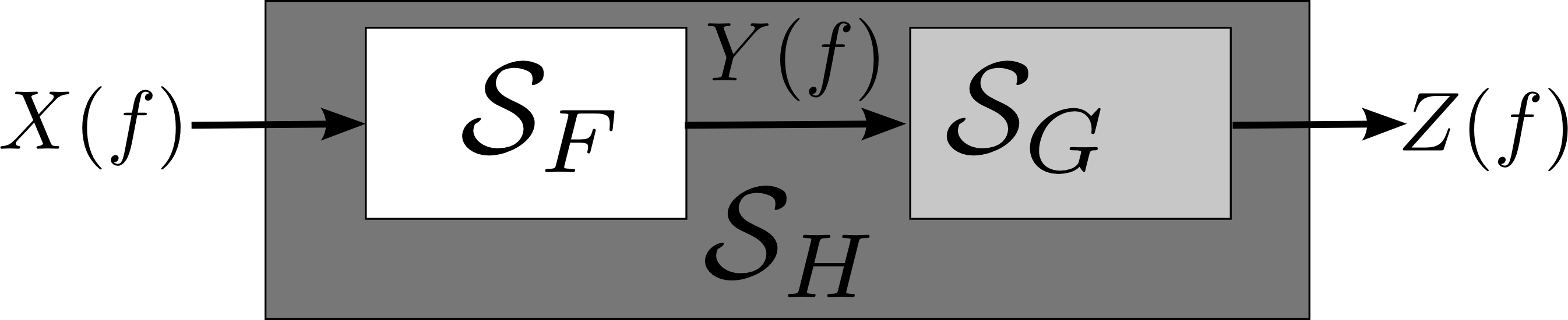}
\caption{A chain of two non-linear systems.}
\label{fig:NL_chain}
\end{center}
\end{figure}

In Fig.~\ref{fig:NL_chain} two non-linear systems modelling the exciting device (an electrodynamic loudspeaker) and the system under study (the soundboard) are chained. The system $\mathcal{S}_F$ transforms its input signal $X(f)$ into $Y(f)$, which becomes the input of the system $\mathcal{S}_G$ and $Z(f)$ denotes the output of the whole chain. $\mathcal{S}_H$ stands for the non-linear system equivalent to the chaining of $\mathcal{S}_F$ and $\mathcal{S}_G$. 

It is assumed here that only $X(f)$ can be directly linearly driven and that both $Y(f)$ and $Z(f)$ are measurable. The contribution of the non-linear system $\mathcal{S}_G$, which cannot be directly linearly driven, has to be estimated.

\subsection{Modelling non-linear systems}

Volterra series are a convenient tool to express analytically the relationship between the input $e(t)$ and the output $s(t)$ of a weakly non-linear system~\cite{Palm1978,Boyd1985} which is fully characterized by the knowledge of its Volterra Kernels in the frequency domain $\{V_k(f_1,{\scriptstyle{\ldots}},f_k)\}_{k \in \mathbb{N}^*}$. 

Cascade of Hammerstein models constitute an interesting subclass of Volterra systems whose Kernels possess the following property:

\begin{equation}
\label{eq:linkVoltHam_freq}
\forall k, \exists \ \tilde{V}_k: \
\forall(f_1,\ldots,f_k), \ V_k(f_1,\ldots,f_k) = \tilde{V}_k(f_1 + \ldots + f_k) 
\end{equation}

Volterra Kernels of cascade of Hammerstein models can thus be expressed as functions of only one frequency variable and are in practice easy to estimate experimentally~\cite{Rebillat2011b,NOV2010}. This simple method is based on a phase property of exponential sine sweeps and the Kernels of such a model can be estimated from only one measured response of the system.

\subsection{Non-linear system equivalent to the chain}

Let the Volterra Kernels $\{F_k(f_1,{\scriptstyle \ldots},f_k)\}_{k \in \mathbb{N}^*}$, $\{G_k(f_1,{\scriptstyle \ldots},f_k)\}_{k \in \mathbb{N}^*}$ and $\{H_k(f_1,{\scriptstyle \ldots}, f_k)\}_{k \in \mathbb{N}^*}$ describe the systems $\mathcal{S}_F$, $\mathcal{S}_G$ et, $\mathcal{S}_H$. In the cascade-case presented in Fig.~\ref{fig:NL_chain}, Volterra Kernels of $\mathcal{S}_H$ can be expressed analytically as functions of the Volterra Kernels of $\mathcal{S}_F$ and $\mathcal{S}_G$ following \cite{Hasler1999}. For $k=1$, one obtains:

\begin{equation}
			H_1(f_1) = \ F_1(f_1) G_1(f_1) 
			\label{eq:NL_H1}
\end{equation}

This proves rigorously the intuitive result that the linear transfer function of a cascade of weakly non-linear systems is the product of the linear transfer functions of each systems composing the cascade. For $k=2$, the following expression is obtained:

\begin{eqnarray}
      H_2(f_1,f_2) = \ F_2(f_1,f_2)G_1(f_1+f_2) + \ldots \nonumber \\
      \qquad \qquad \ldots \quad F_1(f_1)F_1(f_2)G_2(f_1,f_2) 
      \label{eq:NL_H2}
\end{eqnarray}

Assuming that $\mathcal{S}_F$ and $\mathcal{S}_G$ can be modelled as cascade of Hammerstein models, Eq.~(\ref{eq:NL_H2}) becomes:

\begin{eqnarray}
      H_2(f_1,f_2) = \ \tilde{F_2}(f_1+f_2)\tilde{G_1}(f_1+f_2) + \ldots \nonumber \\
      \qquad \qquad \ldots \quad \tilde{F_1}(f_1)\tilde{F_1}(f_2)\tilde{G_2}(f_1+f_2) 
      \label{eq:NL_H2_bis}
\end{eqnarray}

A chain of cascade of Hammerstein models can thus in general not be modelled as a cascade of Hammerstein models as it does not fulfil Eq.~(\ref{eq:linkVoltHam_freq}) for $k=2$ due to the $\tilde{F_1}(f_1)\tilde{F_1}(f_2)$ term.

\subsection{Non-linear contribution of the system $\mathcal{S}_G$}

The output of the system $\mathcal{S}_G$ can generally be decomposed in its linear and non-linear parts as follows:			
	
\begin{equation}
	Z(f) = \tilde{G}_1(f) Y(f) + Z_{NL}^{G}(f)
	\label{eq:def_NL_ES}	
\end{equation}
	
Now suppose that $\mathcal{S}_F$ and $\mathcal{S}_H$ can be modelled as cascade of Hammerstein models and that their Kernels have been estimated using the method proposed in~\cite{Rebillat2011b}. Linear transfer functions $\tilde{F}_1(f)$ and $\tilde{H}_1(f)$ and the signals $Y(f)$ and $Z(f)$ are thus known. The non-linear contribution of $\mathcal{S}_G$ is then:

\begin{equation}
	C_{\mathcal{S}_G}(f) = \frac{Z_{NL}^{G}(f)}{\tilde{G}_1(f) Y(f)} 
\label{eq:apport_NL}
\end{equation}
	
Using, Eqs.~(\ref{eq:NL_H1}) and~(\ref{eq:def_NL_ES}) and multiplying by $\tilde{F}_1(f)$, $C_{\mathcal{S}_G}(f)$ can then be conveniently computed as:

\begin{equation}
	C_{\mathcal{S}_G}(f) = \frac{\tilde{F}_1(f)Z(f) - \tilde{H}_1(f) Y(f)}{\tilde{H}_1(f)Y(f)}
\label{eq:apport_NL_est}
\end{equation}

\subsection{Numerical validation}

The procedure described previously to estimate the non-linear contribution of the system $\mathcal{S}_G$ in the chain of Fig.~\ref{fig:NL_chain} is now validated on a numerical example. 

\begin{figure}[ht!]
\begin{center}
\includegraphics[width = 0.43\textwidth]{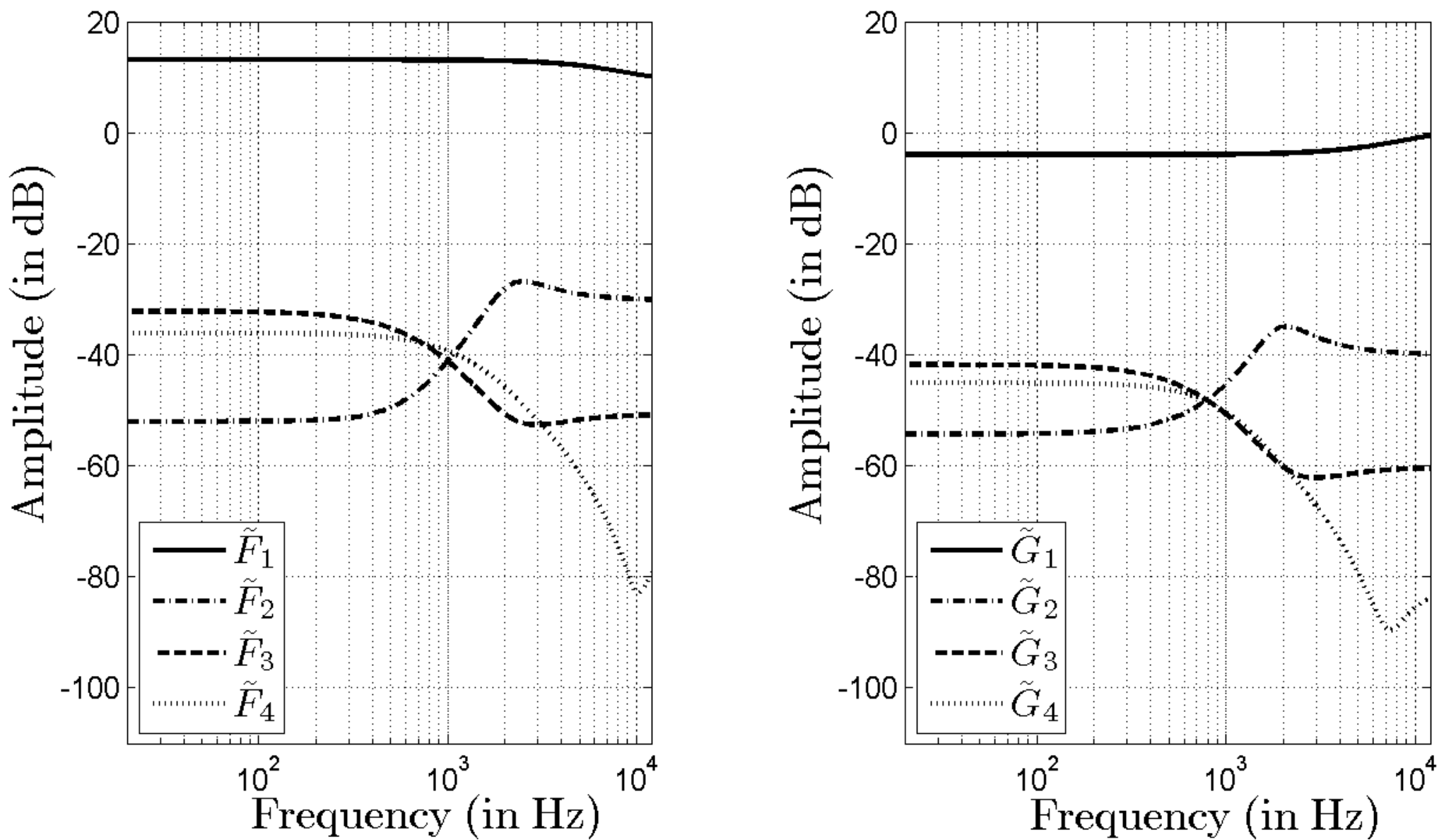}
\caption{Amplitude of the Kernels of the simulated non-linear systems $\mathcal{S}_F$ (left) and of $\mathcal{S}_G$ (right).}
\label{fig:sim_sys}
\end{center}
\end{figure}

Systems $\mathcal{S}_F$ and $\mathcal{S}_G$ have been modelled as cascade of Hammerstein models of order $4$, and each of their Kernels as ARMA filters having two zeros and two poles. The poles and zeros of the different ARMA filters are given in Tab.~\ref{tab:parameters_WNL}. The amplitudes of the Kernels of $\mathcal{S}_F$ and $\mathcal{S}_G$ are presented in Fig.~\ref{fig:sim_sys}. The sampling frequency is chosen as equal to $96$~kHz.

\begin{table}[ht!]
\begin{center}	
\begin{tabular}{|c|l|c|c|c|c|c|}
\hline
  & \multirow{2}{*}{$n$} & $f_{zeros}$ & \multirow{2}{*}{$|p_{zeros}|$} & $f_{poles}$ &  \multirow{2}{*}{$|p_{poles}|$} & \multirow{2}{*}{Gains}  \\
  &     &  \small{(kHz)} &   & \small{(kHz)}  &   &   \\
\hline
\multirow{4}{*}{$\mathcal{S}_f$} & $1$ & $0.15$ &  $0.5$ & $1.5$ & $0.6$ & $3$ \\
																 & $2$ & $0.4 $ &  $0.97$ & $2  $ & $0.95$ & $3\times10^{-2}$ \\
																 & $3$ & $2   $ &  $0.93$ & $0.1$ & $0.95$ & $3\times10^{-3}$ \\
																 & $4$ & $10  $ &  $0.92$ & $0.5$ & $0.92$ & $3\times10^{-4}$ \\
\hline
\multirow{4}{*}{$\mathcal{S}_g$} & $1$ & $0.1$ &  $0.6$ & $1.2$ & $0.5$ & $1$ \\
																 & $2$ & $0.3 $ &  $0.95$ & $1.8  $ & $0.96$ & $10^{-2}$ \\
																 & $3$ & $2   $ &  $0.93$ & $0.12$ & $0.95$ & $10^{-3}$ \\
																 & $4$ & $7  $ &  $0.92$ & $0.5$ & $0.95$ & $10^{-4}$ \\
\hline
\end{tabular}
\caption{Poles and zeros of the ARMA filters used to simulate the non-linear systems $\mathcal{S}_f$ and $\mathcal{S}_g$.}
\label{tab:parameters_WNL}
\end{center}
\end{table}

The Kernels of the non-linear systems $\mathcal{S}_F$ and $\mathcal{S}_H$ have been afterwards estimated using the method presented in \cite{Rebillat2011b} between $20$~Hz and $10$~kHz, with $10$~second exponential sweeps, and assuming non-linear systems of order $4$. 

\begin{figure}[ht!]
\begin{center}
\includegraphics[width = 0.43\textwidth]{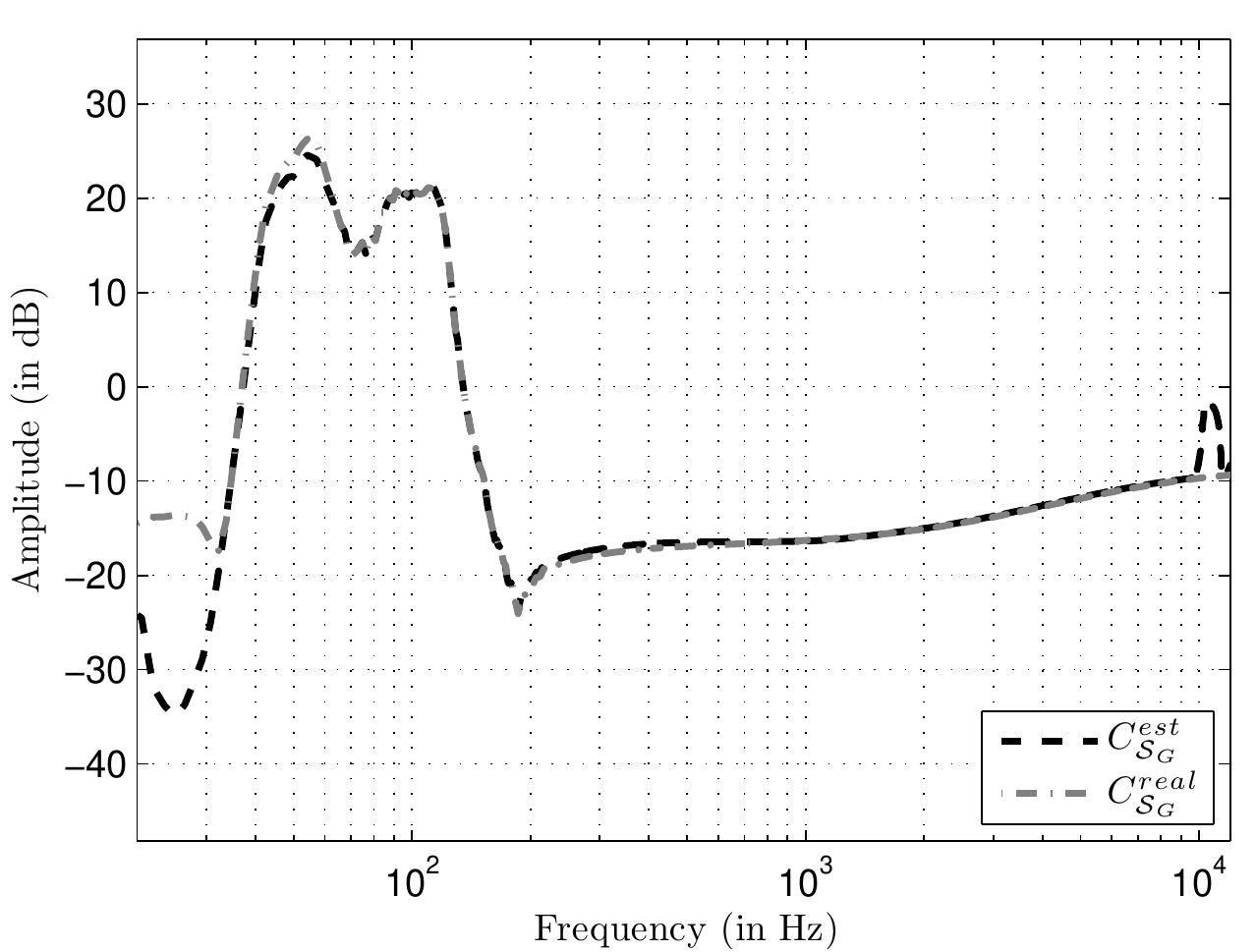}
\caption{$C_{\mathcal{S}_G}^{est}(f)$, the non-linear contribution of $\mathcal{S}_G$ estimated using Eq.~(\ref{eq:apport_NL_est}) and its real value $C_{\mathcal{S}_G}^{real}(f)$ computed from the different Kernels of the non-linear systems.}
\label{fig:NL_est}
\end{center}
\end{figure}

$C_{\mathcal{S}_G}^{est}(f)$, the non-linear contribution of $\mathcal{S}_G$ is then estimated using Eq.~(\ref{eq:apport_NL_est}) and its real value $C_{\mathcal{S}_G}^{real}(f)$ is also computed from the knowledge of the different Kernels of the non-linear systems. $C_{\mathcal{S}_G}^{est}(f)$ and $C_{\mathcal{S}_G}^{real}(f)$ are plotted in Fig.~\ref{fig:NL_est}.

The agreement between the estimated non-linear contribution and the real non-linear contribution is found to be very good even if the system $\mathcal{S}_H$ is approximated by a cascade of Hammerstein models which is mathematically not true here, as shown by Eq.~(\ref{eq:NL_H2_bis}). This thus validate the use of the proposed method to estimate the non-linear contribution of the system $\mathcal{S}_G$ which cannot be directly linearly driven in a chain of two non-linear systems.

\section{Application to string instruments}
\label{sec:results}

\subsection{Experimental protocol}
\begin{figure}[ht!]
\begin{center}
\includegraphics[width = 0.45\textwidth]{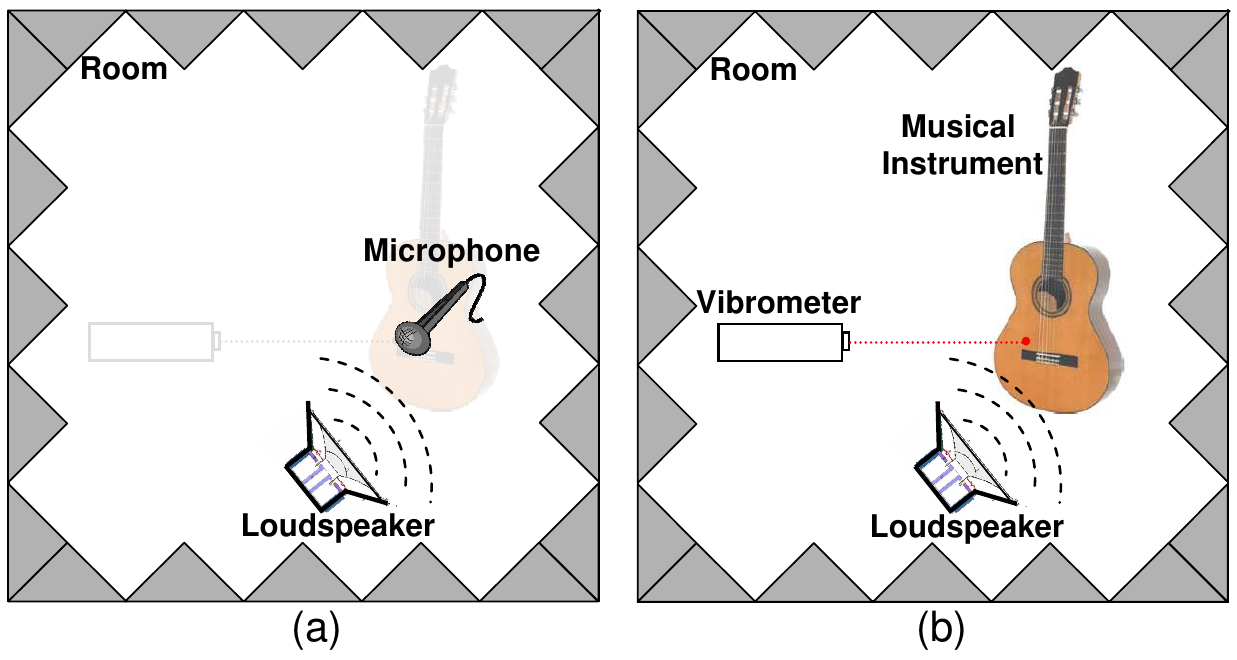}
\caption{Experimental protocol: (a) first configuration (system $\mathcal{S}_F$). (b) second configuration (chain of systems $\mathcal{S}_H$). In the case of the piano, the vibrometer was replaced in the second configuration by accelerometers on the soundboard.}
\label{fig:expe_protocol}
\end{center}
\end{figure}

The soundboard non-linearities of four strings instruments of no particular merit have been estimated: one piano, two guitars and one violin. The experimental protocol (first developed in~\cite{EGE2011}), similar for each of the instruments, is presented hereafter and drawn in Fig.~\ref{fig:expe_protocol}.

\begin{figure}[ht!]
\begin{center}
\includegraphics[width = 0.20\textwidth]{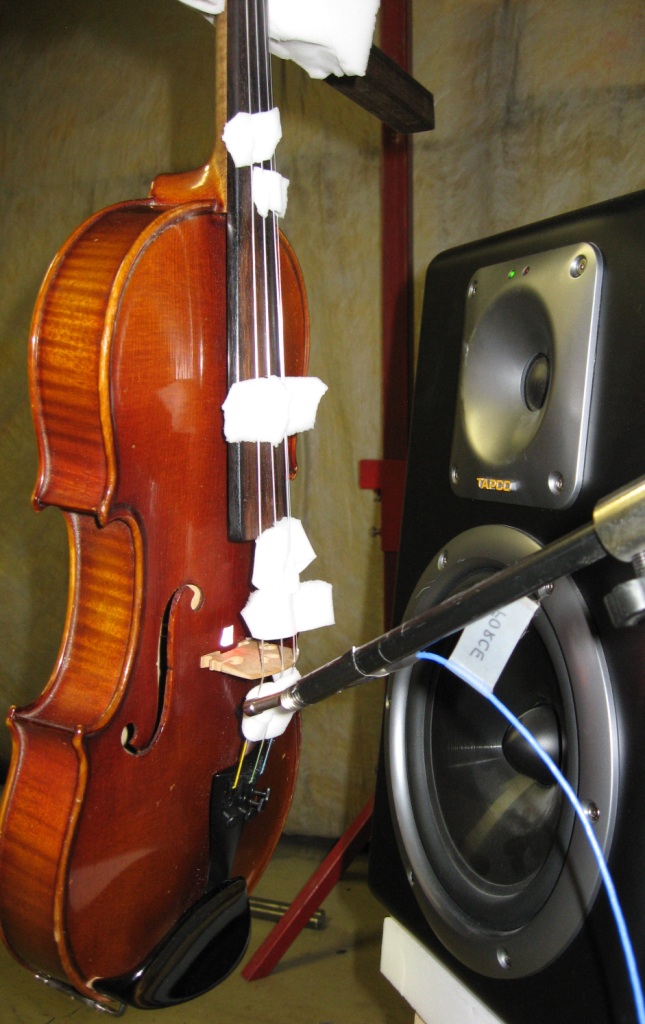}
\caption{Typical measurement on a violin (in playing conditions, with muted strings) excited by a loudspeaker. The reflective adhesive tape (where the velocity is measured by the laser vibrometer) is visible on the violin soundboard, just above the bridge.}
\label{fig:violin}
\end{center}
\end{figure}

The instrument tuned normally and in playing conditions is put in a pseudo-anechoic room (anechoic walls and ceiling, ordinary ground). In the case of the guitars and the violin, the instrument is suspended vertically, clamped at the neck. A particular attention is taken to mute the strings by strips of foam (or woven in two or three places) inserted between them (see Fig.~\ref{fig:violin} for example).  Two configurations \{\emph{loudspeaker, room}\} and \{\emph{loudspeaker, instrument, room}\} have been analysed with the following procedure. The electrical excitation of the loudspeaker was an exponential swept-sine [50-4000]~Hz (26~s duration). For each instrument, the amplitude of the loudspeaker was adjusted at the beginning of the study so to obtain displacements of the soundboard corresponding to realistic playings.  To give an idea, the $G=0.5$ gain (see below), generates soundboards displacements at 500~Hz of around $10^{-5}$~m for the guitars and violin. For the piano soundboard the displacement at 500~Hz is approximately $10^{-6}$~m. According to Askenfelt
and Jansson~\cite{ASK1992} these values correspond to the \emph{ff} playing (see measurements at the bridge of a grand piano reported in Fig.~\ref{fig:Ask_soundb_disp}). 

In the first configuration -- \{\emph{loudspeaker, room}\} -- the acoustic response of the room $y(t)$ is measured with a microphone placed in front of the loudspeaker (where the instrument is to be put in the second configuration). 
In the second configuration -- \{\emph{loudspeaker, instrument, room}\} -- where the instrument replaces the microphone, the motion of the soundboard $z(t)$ was measured with a laser vibrometer for the violin and the guitars (velocity in this case), and with accelerometers for the piano (acceleration in this case). 

Hence, the corresponding input-output scheme for the different configurations can be summarised by the diagram Fig.~\ref{fig:NL_chain}, where each system is weakly non-linear. The first configuration (a) in Fig.~\ref{fig:expe_protocol} corresponds to the non-linear system $\mathcal{S}_F$ and the second one (b) to the chain of two non-linear systems $\mathcal{S}_H$. Hence, the method presented in Sec.~\ref{sec:chain} allows the estimation of the non-linear contribution of the soundboard of the instrument -- second (sub-)system $\mathcal{S}_G$ -- which cannot be directly linearly driven.

For the present purpose, it will be assumed that the instrument response exhibits the same amount of non-linearity in the second configuration as it would if it was excited by the acoustical field that creates $y(t)$ at the microphone. In other words, we consider that the instrument behaves, as far as non-linearities are concerned, like a slightly non-linear (and localised) microphone replacing the true one. 

\subsection{Results}
\begin{figure}[ht!]
\begin{center}
\includegraphics[width = 0.47\textwidth]{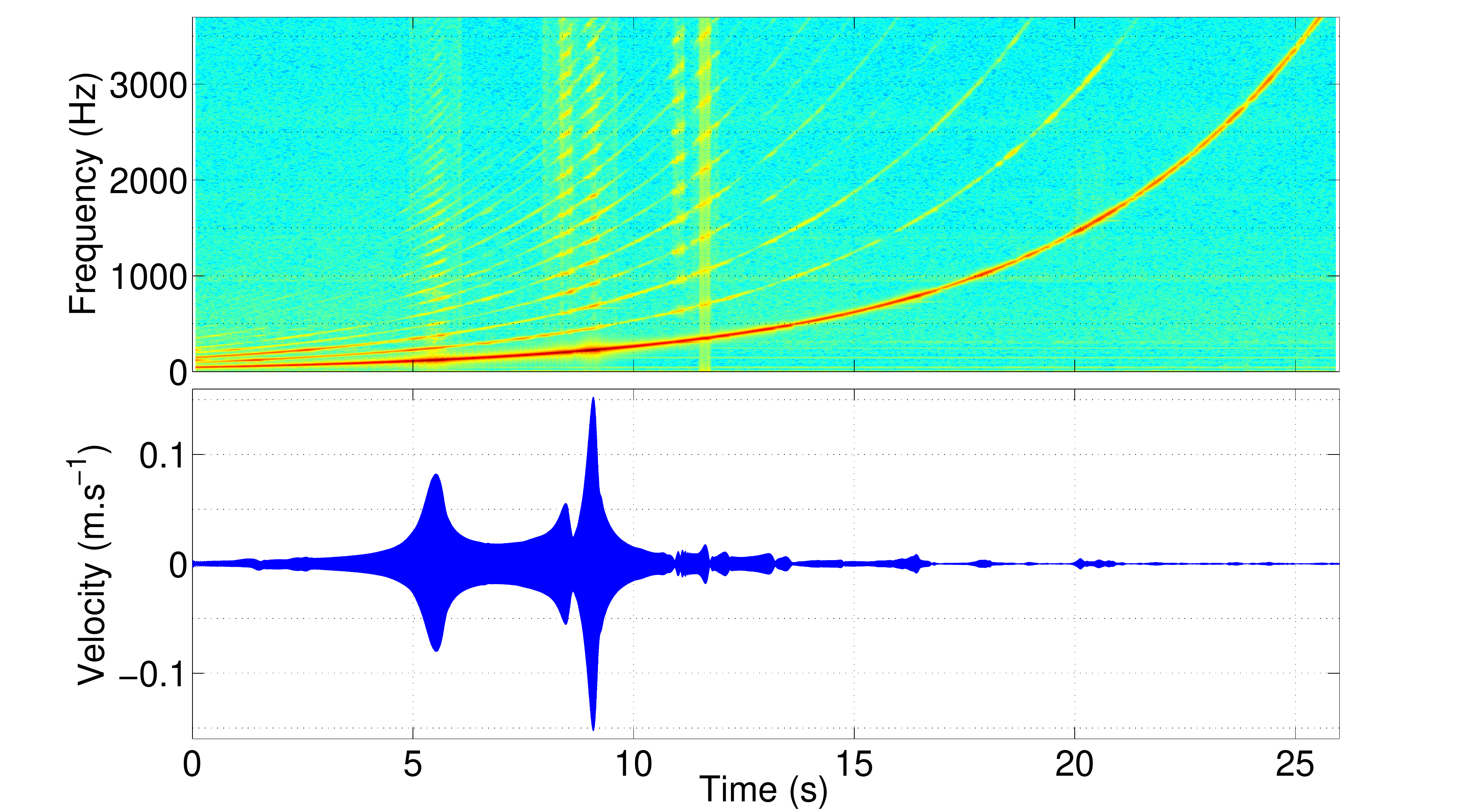}
\caption{Spectrogram and temporal waveform of the soundboard velocity of the first guitar during a typical measurements.}
\label{fig:velocity}
\end{center}
\end{figure}

An example of measurement showing the temporal waveform and the spectrogram of the soundboard velocity measured with the laser vibrometer is given in Fig.~\ref{fig:velocity} (here for the first guitar). Modes of vibrations of the soundboard are clearly visible when instantaneous frequency of the excitation signal approaches the modal frequencies: increase of velocity (visible in the waveform) results in an increase of non-linearities. 

For the four instruments measured, the spectra of the linear parts of the responses and the relative non-linear contributions of the responses separated as explained in Sec.~\ref{sec:chain}, are shown in Fig.~\ref{fig:piano_NL}, Fig.~\ref{fig:guitar1_NL}, Fig.~\ref{fig:guitar2_NL} and Fig.~\ref{fig:violin_NL} (respectively: piano, first guitar, second guitar and violin). For each figures, the upper plots correspond to the Fourier Transform of the linear impulse reponse (typical ``spectrum'') and the lower plots correspond to the separated non-linear contributions (with the same $x$- and $y$-scale to allow proper comparison between the four instruments). The contributions $C_{\mathcal{S}_F}$ of the loudspeaker are given in dotted lines (first two figures), and the contributions $C_{\mathcal{S}_G}$ for the soundboard of the instruments in solid lines. Except for the piano case where only one loudspeaker's gain was tested, measurements have been done with four or five different loudspeaker's amplitude.


\subsubsection{Piano}
\begin{figure}[ht!]
\begin{center}
\includegraphics[width = 0.45\textwidth]{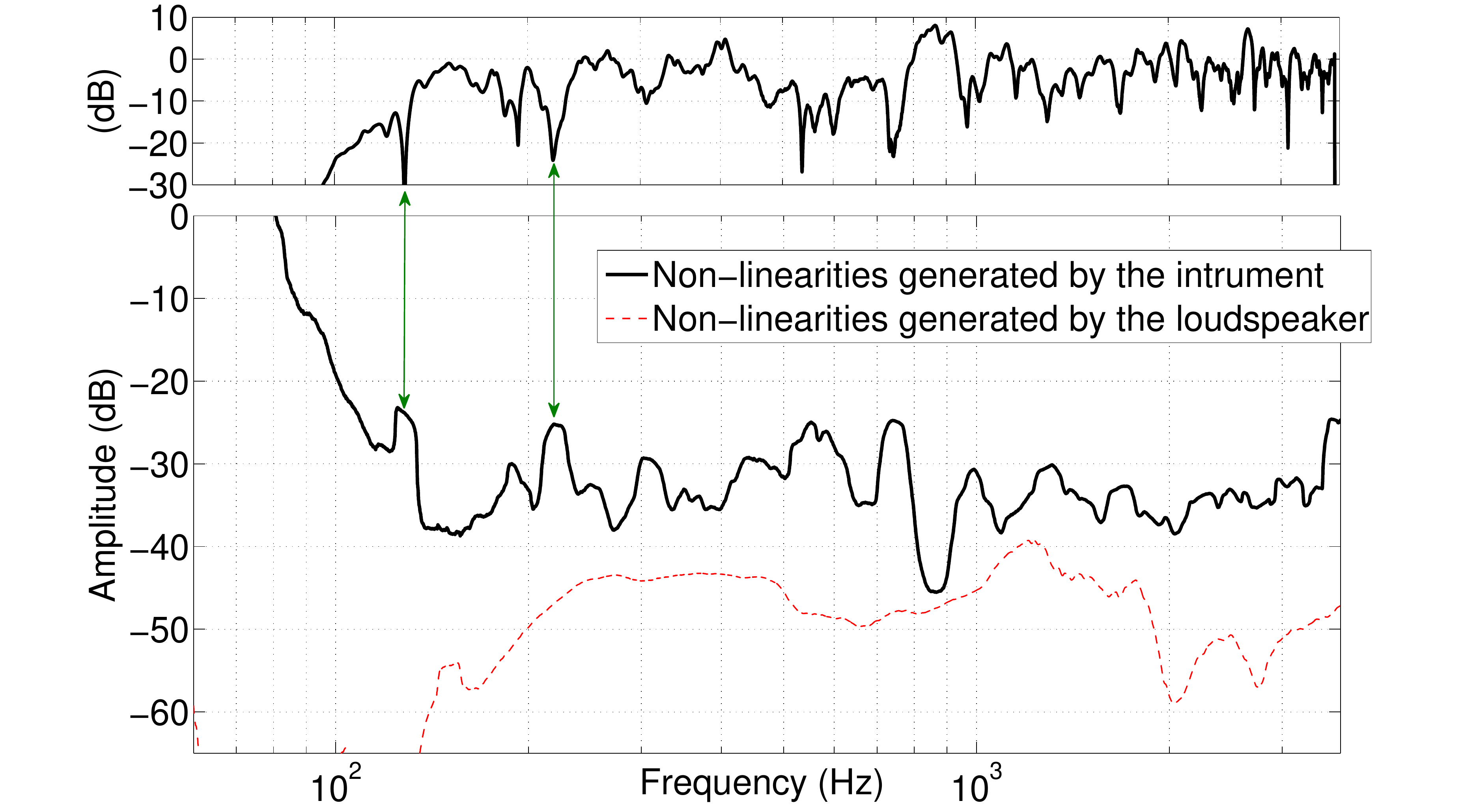}
\caption{Top - Fourier Transform of the linear impulse reponse of the piano soundboard (typical ``spectrum''). Bottom - Measured relative non-linear contributions for both systems $\mathcal{S}_F$ and $\mathcal{S}_G$. Dotted line: non-linearities of the loudspeaker $C_{\mathcal{S}_F}$; Solid line: non-linearities of the piano soundboard $C_{\mathcal{S}_G}$. (Linear contributions are equal to 0~dB on this graph).}
\label{fig:piano_NL}
\end{center}
\end{figure}

The non-linearity content which can be attributed to the piano soundboard appears to be contained within $-30$ to $-50$~dB (see Fig.~\ref{fig:piano_NL}). The apparent increase in non-linearity for frequencies below 100~Hz is probably an artefact of the method since the quality of the reconstruction of the non-linear impulse responses is degraded near the lower and upper bounds of the explored frequency range ([50-4000]~Hz in the present case). Moreover, the increase of non-linearity at anti-resonances (see vertical green double arrows at 129~Hz and 219~Hz for example) is certainly due to the decrease of linear contribution which enhances the measurement noise by construction (see Eqs.~\ref{eq:apport_NL} and \ref{eq:apport_NL_est}). Altogether, the order of magnitude of $-40$~dB can be retained for the non-linear contribution of the piano soundboard at the level of \emph{ff} playing.

Note that in the method proposed in this paper a high quality loudspeaker is essential for an efficient separation of the non-linearity contributions. This quality may be characterised by the amount of ``acoustical'' non-linearity in the first configuration (dotted line in Fig.~\ref{fig:piano_NL}). For the piano measurements the non-linearities of the loudspeaker (Bose - 802 Series II) are contained within $-40$ to $-60$~dB. In the guitars and violin cases (next three figures), the loudspeaker (Tapco S8) has non-linearities contained within $-30$ to $-50$~dB for the $G=0.5$ gain.


\subsubsection{Guitars}
\begin{figure}[ht!]
\begin{center}
\includegraphics[width = 0.45\textwidth]{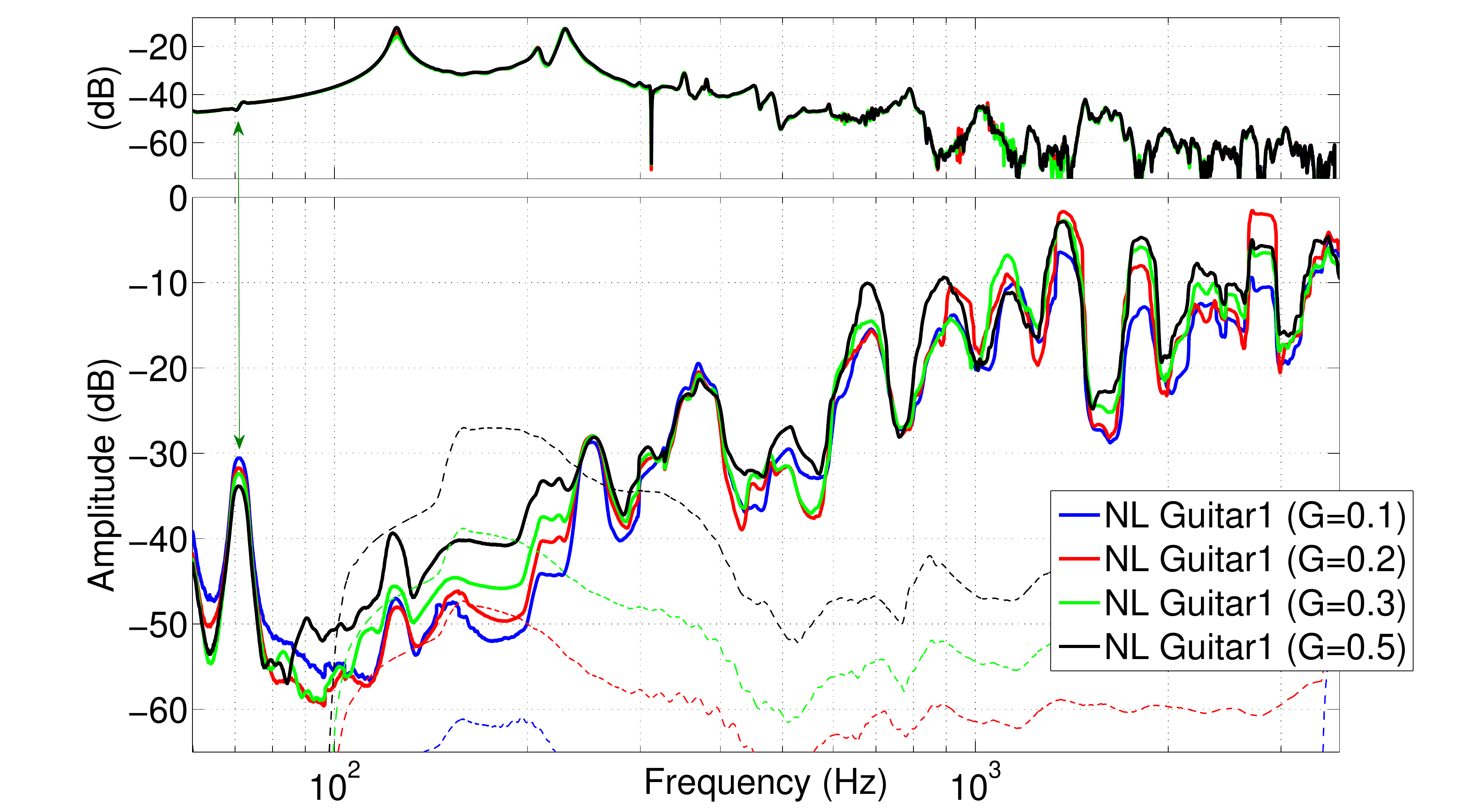}
\caption{Non-linearities generated by the soundboard of the first guitar and estimated for different gain of the loudspeaker. Caption similar to Fig.~\ref{fig:piano_NL}.}
\label{fig:guitar1_NL}
\end{center}
\end{figure}

\begin{figure}[ht!]
\begin{center}
\includegraphics[width = 0.45\textwidth]{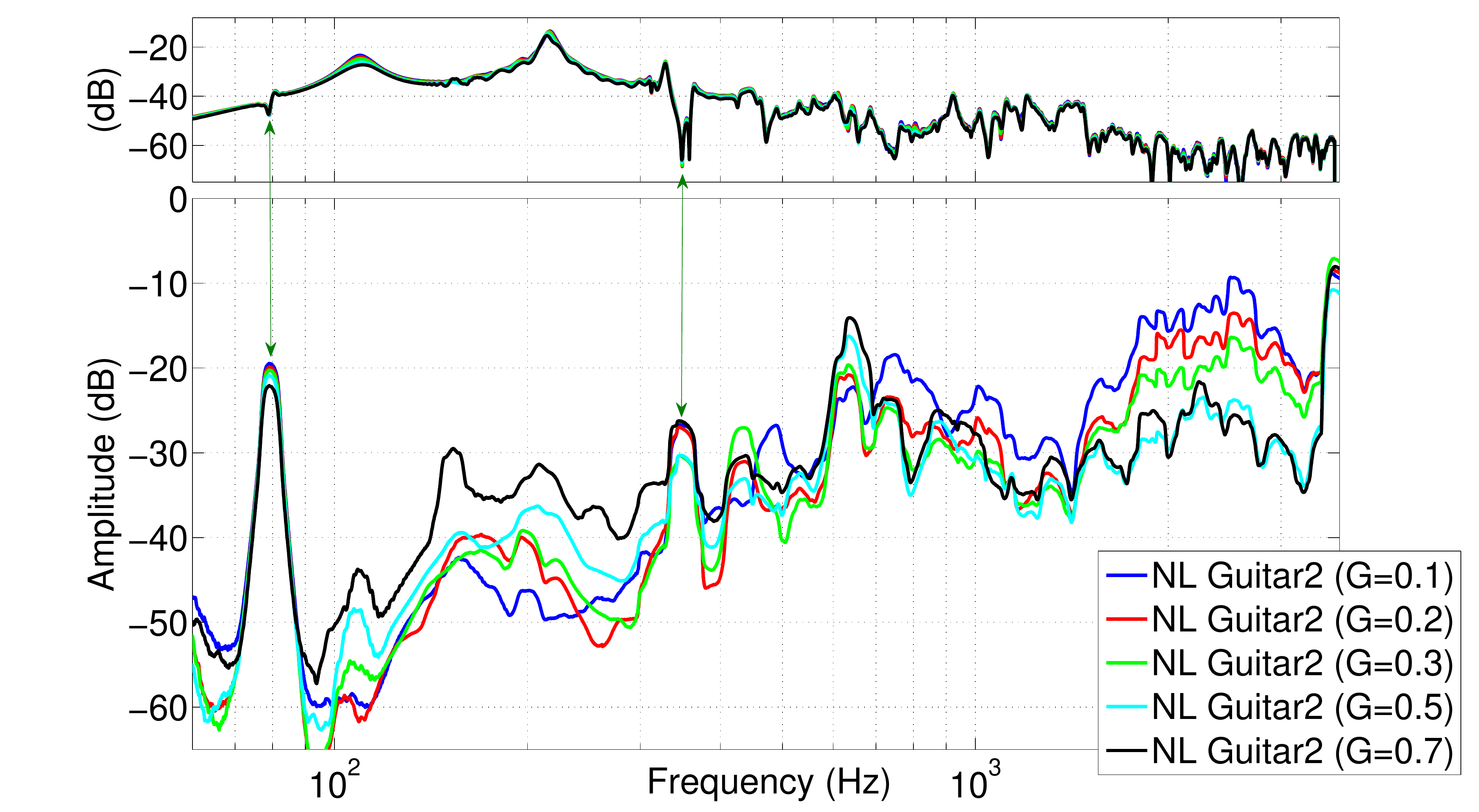}
\caption{Non-linearities generated by the soundboard of the second guitar and estimated for different gain of the loudspeaker. Caption similar to Fig.~\ref{fig:piano_NL}. (Dotted lines are removed for more clarity.)}
\label{fig:guitar2_NL}
\end{center}
\end{figure}

The results on the two guitars are given in Figs.~\ref{fig:guitar1_NL} and~\ref{fig:guitar2_NL} for several gain $G$ of the loudspeaker. The spectra of the linear impulse responses are independent of the gain (almost undistinguishable one from the other; see the top of the figures) which is consistent with the theory.

As explained above, some peaks of non-linearity are clearly visible at anti-resonances, specially before the Helmholtz mode (first resonance) of the guitars (see green double arrows at 71~Hz for the first instrument and 79~Hz for the second one). 

In the low and mid-frequency ranges the evolution with frequency of the non-linearities are similar for both guitars: a constant increase from -50~dB (at 100~Hz) to -30~dB at 600~Hz. For frequencies higher, the non-linearities of the second guitar seems to reach a plateau level of -20~dB (for the two highest excitation level) whereas for the first guitar this increase continues and reaches -10~dB at 3.5~kHz.

Moreover the non-linearities of the soundboards increase with $G$ for both guitars, particularly in the frequency band [100-700]~Hz which is also consistent with the theory. The fact that in the high frequencies this evolution of non-linearities with the gain is inverted for the second guitar may still be attributed to an increase of the signal to noise ratio (SNR) which leads to a decrease of the artefacts caused by noise (when G increases). This explains also why at anti-resonances the lower gains (lower SNR) increases the amount of estimated non-linearities (see for example the blue curves higher than black ones at anti-resonance of the Helmholtz modes, for both guitars).

\subsubsection{Violin}
\begin{figure}[ht!]
\begin{center}
\includegraphics[width = 0.45\textwidth]{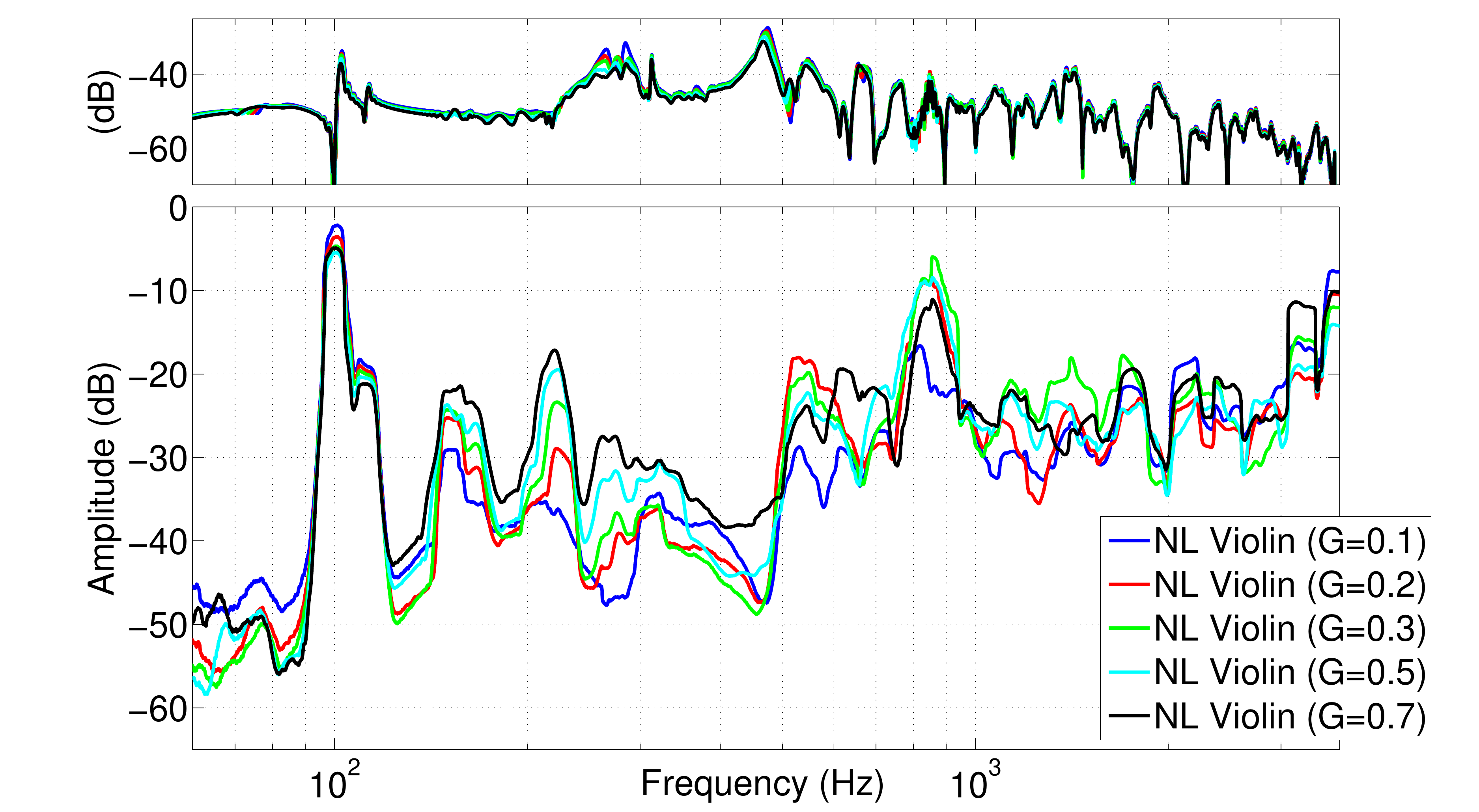}
\caption{Non-linearities generated by the soundboard of the violin and estimated for different gain of the loudspeaker. Caption similar to Fig.~\ref{fig:piano_NL}. (Dotted lines are removed for more clarity.)}
\label{fig:violin_NL}
\end{center}
\end{figure}

Results on the violin are given in Fig.~\ref{fig:violin_NL}. Except for the peaks at anti-resonances the violin soundboard intrinsic non-linearity is contained within -20~dB to -50~dB.

\subsection{Discussion}
\begin{figure}[ht!]
\begin{center}
\includegraphics[width = 0.47\textwidth]{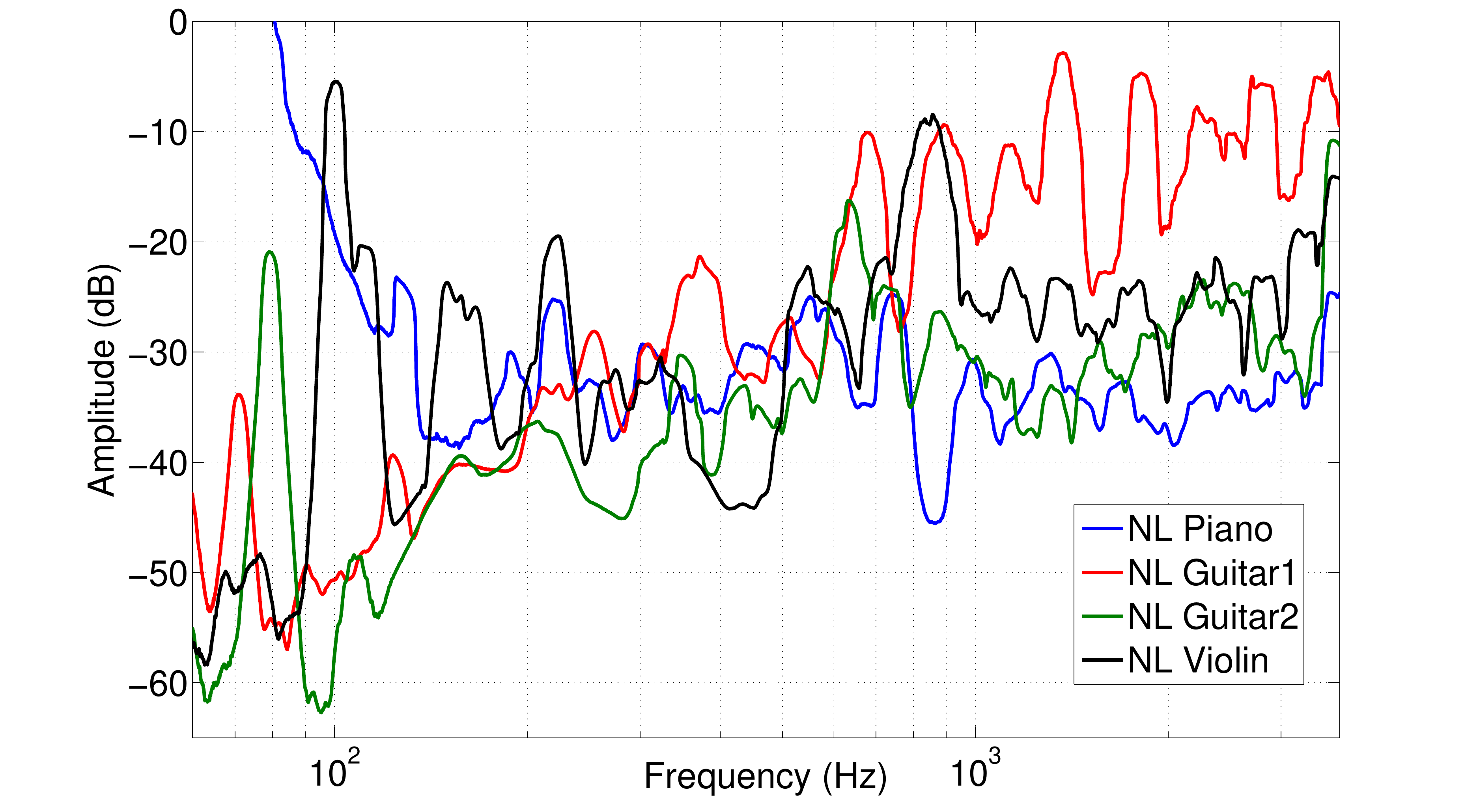}
\caption{Comparison of the non-linearities generated by the four instruments. ($G=0.5$ for the guitars and the violin.)}
\label{fig:4instru}
\end{center}
\end{figure}

The assumption of linearity of the soundboard vibrations of the four instruments is verified to a high level of approximation (see Fig.~\ref{fig:4instru} where the results are plotted in the same graphs). 
The mean value of non-linearities is more than 20~dB less than the linear parts in all the cases, except near the lower and upper bounds of the explored frequency range ([50-4000]~Hz) (artefact of the method) or at anti-resonances where the SNR decreases. The comparison reveals that the non-linearity content of the first guitar (red line) is almost 10~dB more than for the other instruments, for frequencies higher than 1~kHz. A conclusion on the quality of this instrument is unfortunately impossible here (and is not the object of this paper). Such conclusions would require measurements with more instruments of different qualities/origins...

\section{Conclusion \& Perspectives}
In this article an original vibro-acoustical method is presented to isolate the soundboard non-linearity of string instruments from that of the exciting device (here a loudspeaker) and to measure it. For a chain of two non-linear systems, the method allows the estimation of the non-linear contribution of the second system which cannot be directly linearly driven. Experimental quantifications of the linear approximation of the intrinsic soundboard vibrations of one upright piano, two guitars and one violin is given for level of excitation corresponding to the \emph{ff} playing. These non-linearities are contained within 20~dB to 50~dB less than the linear parts in the [50-4000]~Hz frequency range, except in the high frequency domain of one of the guitars. The measurement noise appears to be crucial for a proper estimation of these non-linearities. A technique which allows to separate the measurement noise from the system non-linearity and to measure it (as in \cite{ZHA2010}) must be integrated in the method before giving conclusions in terms of musical quality of the measured instruments.
\section*{Acknowledgments}
The authors would like to thank LVA PhD Students -- Rainer Stelzer and Roch Scherrer -- for the lending of their instruments.
\bibliographystyle{unsrt}
\bibliography{NL_bib}

\end{document}